\documentclass[prl,twocolumn,showpacs,preprintnumbers,amsmath,amssymb]{revtex4-1}

\usepackage{graphicx,natbib}
\usepackage{dcolumn}
\usepackage{bm}

\begin{document}

\title{Emergence of order from turbulence in an isolated planar superfluid}

\author{Tapio Simula$^1$, Matthew J. Davis$^2$, Kristian Helmerson$^1$}
\affiliation{$^1$School of Physics, Monash University, Victoria 3800, Australia}
\affiliation{$^2$School of Mathematics and Physics, University of Queensland, Queensland 4072, Australia}

\begin{abstract}
{We study the relaxation dynamics of an isolated zero temperature quasi-two-dimensional superfluid Bose-Einstein condensate (BEC) that is imprinted with a spatially random distribution of quantum vortices.  Following a period of vortex annihilation, we find that the remaining vortices self-organise into two macroscopic coherent `Onsager vortex' clusters  that are stable indefinitely.  We demonstrate that this occurs due to a novel physical mechanism --- the evaporative heating of the vortices --- that results in a negative temperature phase transition in the vortex degrees of freedom.   At the end of our simulations the system is trapped in a non-thermal state.  Our computational results provide a pathway to observing Onsager vortex states  in a superfluid Bose gas.}
\end{abstract}

\maketitle

The question of how thermodynamics arises from unitary quantum evolution \cite{Mahler} has been debated since the early days of quantum mechanics. The closest laboratory realisation of an isolated quantum system of many particles is perhaps the ultracold quantum gas, and recent progress in the control and manipulation of these systems mean that the question is no longer simply an academic one \cite{Polkovnikov2011a}.
A particular focus has been one-dimensional Bose gases as described by the Lieb-Liniger model~\cite{LiebLiniger}, as the integrability of the model suggests that it may be prevented from attaining thermal equilibrium following a quench~\cite{Rigol2009a}.  Indeed, two groundbreaking experiments on the dynamics of one-dimensional Bose gases~\cite{Weiss2006-Nature,Schmiedmayer2007-Nature} sparked a rush of further activity due to their seemingly contradictory results on whether the experiments returned to standard thermal equilibrium. These experiments have generated significant recent theoretical interest in the nonequilibrium dynamics and relaxation of idealised isolated quantum systems following a disturbance~\cite{Rigol2008-Nature,Polkovnikov2011a,Dziarmaga2010a,Djunko2012a}. 

The quantum relaxation of higher dimensional isolated quantum systems is difficult to address computationally due to their exponential complexity. Recent experiments have quenched the interatomic interaction strength of 3D BECs in oblate and spherical harmonic traps and followed the subsequent dynamics ~\cite{Makotyn2013,Hung2013}, including apparent saturation of the momentum distribution ~\cite{Makotyn2013}. However the relevance of quantum dynamics in their relaxation is not clear. In some situations the classical field approximation~\cite{Blakie2008b,Brewczyk2007a} can provide insight to quench dynamics.  Indeed, recent theoretical work has considered the classical equilibration dynamics of 2D superfluids following a quench from the perspectives of turbulence and non-thermal fixed points~\cite{Mathey2010a,Nowak2011a,Nowak2012a,Schole2012a,Shukla2013a,Hofmann2014a}.  

The thermalisation of isolated \emph{classical} systems is generally understood in terms of ergodicity and chaotic dynamics~\cite{Gallavotti1999}.  Even though the equations of motion of a physical system are entirely reversible, if a system with a sufficiently large number of degrees of freedom begins in an `atypical' state --- such as a gas with all particles in one half of the container --- it will  quickly relax to a more `typical' state consistent with thermal equilibrium as predicted by statistical mechanics.  This agrees with our everyday experience of the arrow of time that makes it abundantly clear that time only proceeds in one direction~\cite{Lebowitz1993a}, and is  encapsulated by the second law of thermodynamics --- that isolated systems only become more disordered with time~\cite{Thomson1852}. Here we present results on an isolated, quasi-two dimensional superfluid Bose gas in which order appears with time.

\begin{figure*}
\centering
\includegraphics[width=1.8\columnwidth]{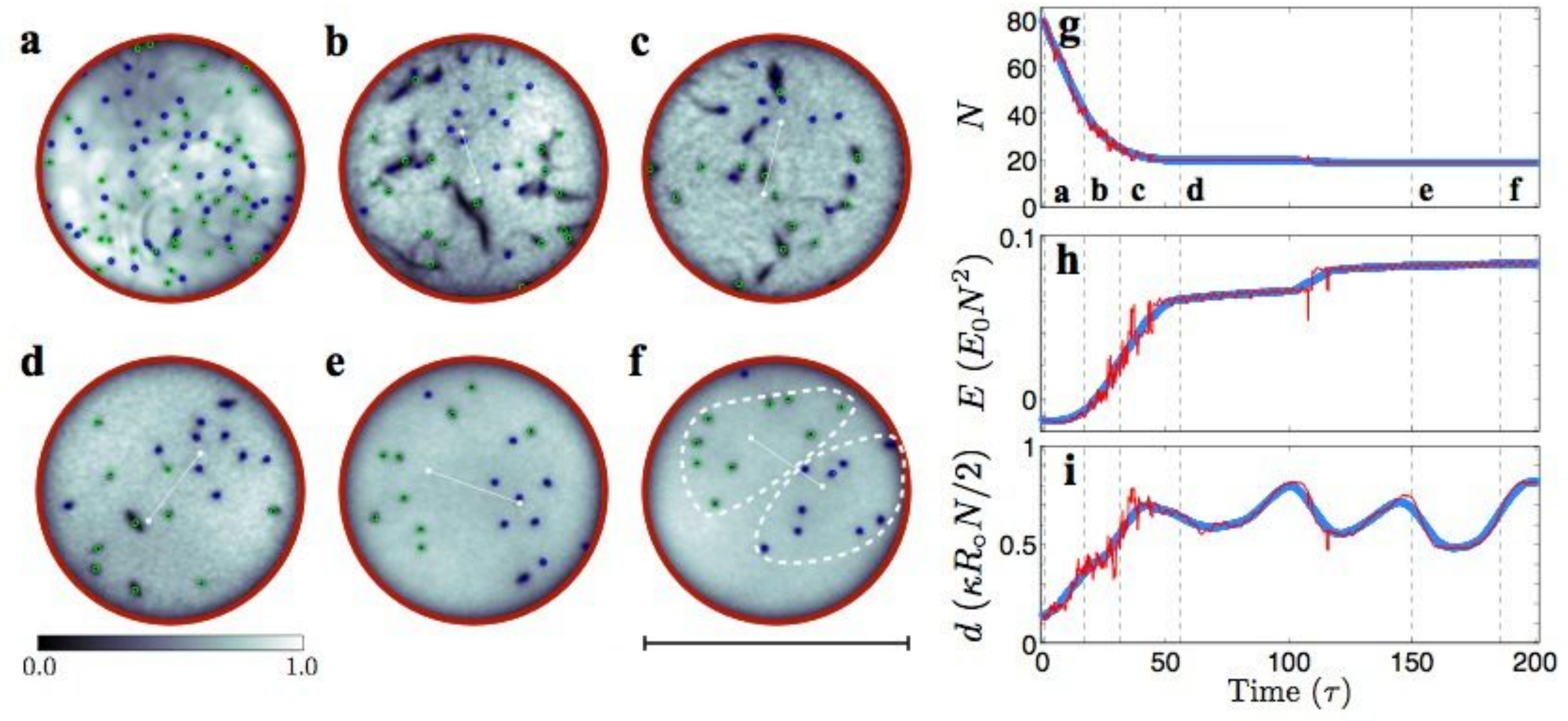}
\caption{Onsager vortex formation in the Gross--Pitaevskii model.  (a--f) Time series of the condensate column density. The location of vortices with positive and negative circulations are shown using blue and green circles, respectively, and the white line indicates the vortex dipole moment vector $\mathbf{d}$ as defined in the text.  The initial random vortex configuration in (a) evolves to the Onsager vortices configuration in (f) characterised by two large-scale, coherent clusters of vortices in a background field of sound. (g) The total number of vortices, $N$, as a function of time, with the dashed vertical lines indicating the times at which we plot the condensate density in (a--f).  (h) Kinetic energy, $E$, of the system. (i) Dipole moment, $d$, of the vortex configuration as a function of time. In (g--i) the red curves correspond to instantaneous values and the smooth thick curves are sliding averages over a window of $15 \tau$ to smooth out rapid fluctuations. The bar under (d) shows the scale of condensate density and the length of the scale bar under (f) is $56\; a_{\rm osc}$, the diameter of the trap.\label{fig1}}
\end{figure*}

Typically quenches in isolated systems increase the energy per particle, leading to more disorder and increased entropy at equilibrium. However, for systems with a limited phase space, continuing to add energy will eventually render it \emph{more ordered}.  This decrease in entropy with increasing energy is the definition of a state  with \emph{negative} absolute temperature~\cite{Onsager1949a,Ramsey1956a}. Such states are actually ``hotter'' than those at positive temperature, as energy will spontaneously flow from negative to positive temperature systems when in contact.  Thus for negative temperature thermodynamic states to be realised in practice they need to be isolated from their environment. Negative temperatures have been realised in spin-systems \cite{Purcell1951a}, and more recently with ultracold atoms in optical lattices~\cite{Braun04012013}. 

Onsager predicted that negative temperature states may be relevant for two-dimensional (2D) fluids by applying statistical mechanics to a 2D model of point vortices~\cite{Onsager1949a}.  The point vortex model represents the full velocity field of the fluid as the superposition of the circular velocity fields generated by the individual vortices. Hence the point vortices themselves have no inertial kinetic energy --- resulting in a phase space determined entirely by the finite area available to the point vortices, thus allowing negative temperatures. While intended as a model of 2D fluids in general, Onsager noted that the model was potentially particularly relevant for 2D superfluids, whose vortices have quantized circulation~\cite{Onsager1949a,Eyink2006a}.  

Experiments on quasi-two-dimensional ultracold quantum gases are now routine~\cite{Hadzibabic2006a,Clade2009a,Kwon2014a}, and offer tantalising prospects for the study of 2D turbulence in superfluids. Recently, Neely \emph{et al.} have demonstrated that coherently stirring a quasi-2D Bose-Einstein condensate (BEC) can lead to a proliferation of vortices, with evidence of transient, local clusters of like-signed vortices~\cite{Neely2012a}. So far, however, a mechanism that enables the observation of Onsager's negative absolute temperature states corresponding to long-lived, giant vortex clusters in inviscid quantum fluids has not been discovered.

In this paper we study the relaxation dynamics of an isolated quasi-two-dimensional superfluid Bose--Einstein condensate (BEC) that is ``quenched'' by imprinting a spatially random arrangement of vortices.  We subsequently simulate the classical conservative Hamiltonian dynamics, and find that two Onsager vortices corresponding to negative temperature states can arise out of the initially turbulent flow.  This emergence of order in an isolated system occurs due to evaporation of vortices that enable a  redistribution of the energy amongst the degrees of freedom, and leaves the BEC in a non-thermal state.

Our simulations begin with a  BEC  at zero temperature, in which we prepare  an equal number of vortices and antivortices (80 in total) at stochastically sampled locations. We subsequently simulate the dynamics of the BEC using the Gross-Pitaevskii equation (GPE), a nonlinear Schr\"odinger equation providing a realistic mean-field description of BECs~\cite{pethicksmith}. The conservative Hamiltonian evolution of the GPE preserves the total energy $E_{\rm GP}$, atom number $N_a$, and angular momentum of the system.  While the BEC itself is three-dimensional, the dynamics of the vortices are essentially two-dimensional \cite{Neely2012a,Rooney2011}. In contrast to previous theoretical studies of quasi-two dimensional quantum turbulent systems in harmonic traps \cite{Parker2005a,Horng2009a} or in uniform doubly-periodic domains \cite{Numasato2010a,Nowak2012a,Reeves2013a,Billam2013a}, we consider a BEC confined in a disc trap. Details of these simulations are provided in the Supplemental Material \cite{SupplementPRL}.

The results of our simulation for a typical initial condition can be seen in Movie S1 \cite{SupplementPRL}. In Fig.~\ref{fig1}(a--f) we show the condensate column densities for a range of evolution times from the movie, and identify the location of the vortices and antivortices. The  dynamics initially leads to the annihilation of several vortex--antivortex pairs, with the subsequent emission  of sound waves in the bulk BEC. Occasionally this process is reversed.  The vortex pair annihilation results in the rapid decay of the total vortex number until  $\sim20$ remain [Fig.~\ref{fig1}(g)], while the GPE kinetic energy per vortex grows [Fig.~\ref{fig1}(h)]. During this annihilation period, like-signed vortices exhibit a tendency to form transient clusters that grow larger with time.  A remarkable feature of this system is the eventual  emergence of two, counter-rotating clusters of like-signed vortices on opposite sides of the condensate, which we identify as Onsager vortices (OVs). These are characterised in Fig.~\ref{fig1}(i) by the growth in the vortex dipole moment 
$d = |{\bf d} |=|\sum_i q_i{\bf r}_i  |$, 
where  ${\bf r}_i $ is the position of the $i$th vortex. The vortex charge $q_i = s_i h/m$, where $s_i=\pm1$ for vortices and antivortices respectively, $h$ is Planck's constant and $m$ is the mass of an atom. After the formation of OVs, the vortex annihilation mostly ceases and the OVs remain a robust and long-lived feature of the system.  Analyses of the GPE energetics and spectral features are provided in the Supplemental Material \cite{SupplementPRL}.

The abrupt end to the annihilation of vortices and the spontaneous emergence and persistence of the OVs in Fig.~\ref{fig1} is a surprising result.   How has order spontaneously arisen from an initially chaotic state?  The vortex gas is apparently trapped in a negative temperature state, whereas the sound waves on top of the BEC have a positive temperature, suggesting that ergodicity has been broken in this system. 

\begin{figure}
\centering
\includegraphics[width=0.9\columnwidth]{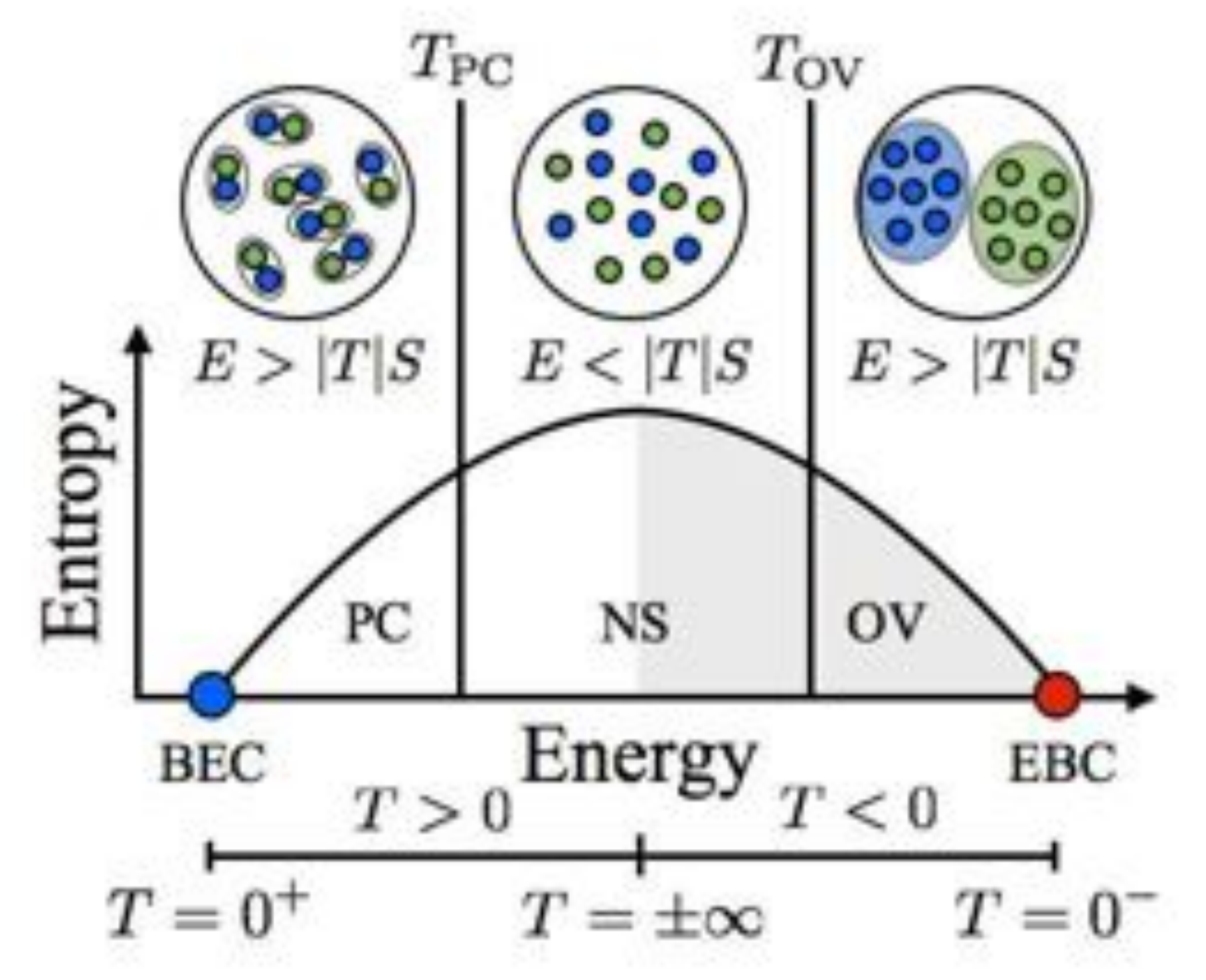}
\caption{A schematic plot of entropy versus energy for the point vortex model. A zero-entropy Bose--Einstein condensate (BEC) forms at $T=0^+$ with its negative temperature counterpart, an Einstein--Bose condensate (EBC) \cite{Kraichnan1967a}, emerging at $T=0^-$. Entropy is maximised at $T=\pm \infty$ in the entropy dominated normal state (NS), which has a stochastic distribution of vortices. The vortex binding-unbinding phase transition separates the normal state from the pair collapse (PC) state at positive temperature, whereas there is a transition to the coherent Onsager vortex (OV) state at a vortex number dependent negative temperature.\label{fig2}}
\end{figure}

In order to develop a microscopic understanding of our results, we utilise Onsager's point vortex model for a two-dimensional fluid, which can be derived as an approximation to the GPE in the incompressible limit~\cite{Lin1999}.   We consider $N$ singly quantized point vortices with equal numbers of each circulation confined in a disk geometry \cite{PointinLundgren1976a,Viecelli1995a,Yatsuyanagi2005a}. In the Supplemental Material \cite{SupplementPRL} we present the results of Monte Carlo calculations for the equilibrium thermodynamics of this system~\cite{Viecelli1995a}. This allows the construction of the schematic phase diagram depicted in Fig.~\ref{fig2}. 

At positive zero temperature ($T=0^+$) in Fig.~\ref{fig2}, a zero entropy, zero momentum Bose--Einstein condensate (BEC) exists.  In the pair-collapse (PC) phase at low positive temperatures,  all vortices and antivortices are paired, and the velocity fields generated by each vortex cancel \cite{Hauge1971a,Kraichnan1980a,Viecelli1995a}. On increasing the temperature there is a critical point where the vortex pairs unbind and the system transitions to the normal state (NS). This is the entropy-dominated  regime where  correlations between the positions of vortices and antivortices are negligible.

The point vortex model has a finite phase space, and the entropy $S$ is not a monotonically increasing function of energy $E$. This allows $\beta = 1/k_{\rm B}T=(\partial S/\partial E) <0$, where $k_{\rm B}$ is the Boltzmann constant, and hence negative temperature states \cite{Onsager1949a,Ramsey1956a}.  Continuing to increase the energy results in another abrupt change in the configuration of the vortices at a critical negative temperature $T_{\rm OV}$, where well-defined clusters of like-signed vortices emerge --- the Onsager vortices.  Eventually, at $T=0^-$ (see Fig.~\ref{fig2}) the OV state corresponds to a pure, zero entropy Einstein--Bose condensate (EBC) --- a BEC in a nonzero momentum state \cite{Kraichnan1967a,Kraichnan1980a}.

Our GPE simulation exhibits counter-intuitive collective behaviour --- an apparent dynamical transition in the vortex gas from an initially  disordered NS phase  to a ordered OV phase at negative temperature.
This result raises the question  --- what is the physical mechanism underlying this emergent phenomenon? In short, the answer is the \emph{evaporative} heating of the vortex gas.  

The kinetic energy of the system can be divided into three components --- an incompressible part due to the rotational vortex velocity field,  a compressible part due to sound (phonon excitations), and a quantum pressure term~\cite{pethicksmith}.   The incompressible kinetic energy in the GPE initial state is significant, while the compressible kinetic energy is small.  The GPE dynamics convert the rotational energy of vortices to sound energy via vortex-antivortex annihilation.  Once the Onsager vortices form, the energy of these two components are approximately constant as shown in Fig.~S3.

The vortex pair annihilation, however, leads to an increase in the mean energy per vortex. The annihilation always occurs  at the length scale of the vortex core size, which is much smaller than the mean distance between vortices. Since the  energy of a vortex-antivortex pair decreases with separation distance~\cite{pethicksmith},  pair annihilation causes a decrease in the total energy of the vortex gas that is small in comparison to the mean energy per vortex, while also reducing its entropy.  The subsequent dynamics of the vortex gas lead to rethermalisation with a larger mean energy per vortex.

This process is \emph{evaporative heating} --- the reverse analog of forced evaporative cooling used to achieve quantum degeneracy in ultracold atomic gases. Evaporative cooling removes  the ``hottest'' atoms from the system, leaving the remaining atoms to rethermalise to a lower positive temperature \cite{Ketterle1996a}. Instead, in our simulation the ``coldest" vortices, corresponding to vortex-antivortex pairs, are removed, leaving the remaining vortices to equilibrate at a higher average energy, and hence a hotter negative temperature. The evaporative heating of the vortex gas becomes ineffective once the Onsager vortices have formed, and encounters between vortices and antivortices become rare.

The evaporative heating mechanism is strikingly confirmed by  dynamical simulations of  Onsager's point-vortex model with the addition of  vortex-antivortex annihilation.  Whenever a pair of vortices of opposite sign come within the ``annihilation distance" $d_a$ of one another, they are removed from the simulation as described in the Supplemental Material \cite{SupplementPRL}. Schwarz used a similar dynamical rule to account for reconnection events in a vortex filament model of three-dimensional superfluid turbulence \cite{Schwarz1988a} and Campbell and O'Neil used forced vortex annihilation to describe viscosity in a two-dimensional vortex random walk model \cite{Campbell1991a}.

\begin{figure}
\centering
\includegraphics[width=0.9\columnwidth]{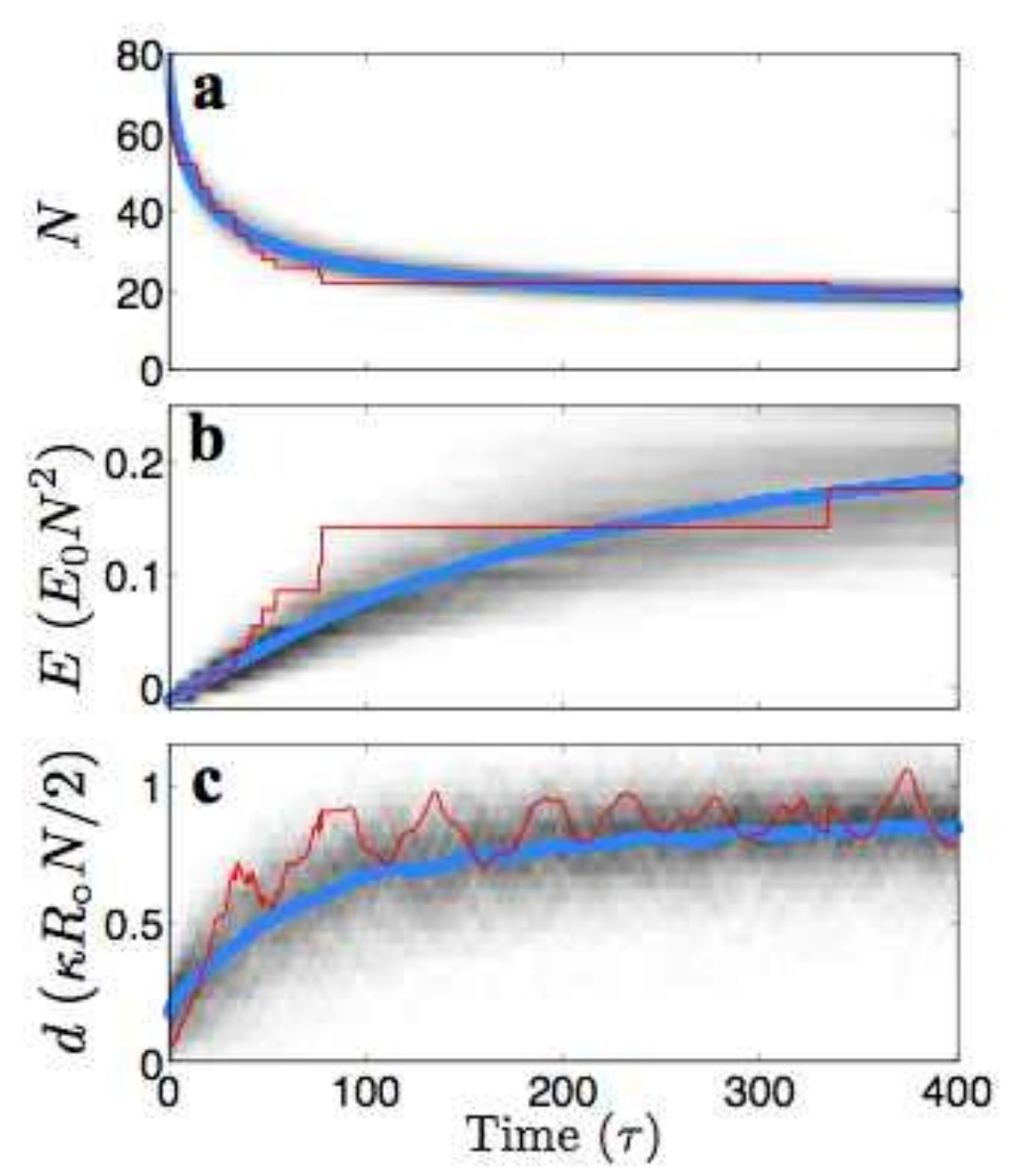}
\caption{Point vortex dynamics with the addition of vortex pair annihilation. (a) Vortex number, $N$, (b) Energy, $E$, and (c) Dipole moment, $d$, as functions of time for an ensemble of 500 simulations with an initial energy corresponding to that of the GPE simulation. Thick smooth lines are ensemble averages, red curves are results from a typical single trajectory, and the grey scale densities are normalised histograms of the simulation ensemble. The slow undulations of the dipole moment in (c)  are due to the large-scale orbital motion of the Onsager vortices.\label{fig3}}
\end{figure}

We have performed simulations for an ensemble of 500 stochastically sampled vortex configurations matched to the GPE initial state, and a sample trajectory is shown in Movie S2 \cite{SupplementPRL}.
The results for the vortex number, energy, and  dipole moment are shown in Fig.~\ref{fig3}(a--c), and should be compared with the corresponding quantities for the GPE simulation in Fig.~\ref{fig1}(g--i). These results confirm that the energy per vortex increases as the vortex pairs evaporate, and this is accompanied by an increase in the net vortex dipole moment, for which  our Monte-Carlo calculations indicate the OV transition occurs near $d\sim 0.5$.  

We hence conclude that the GPE dynamics describe a transition to a coherent negative temperature Onsager vortex state in a closed system --- order emerges from chaos.  This may leave the impression that the arrow of time for the vortices has been reversed.  However,  the heating of the background BEC, through the transfer of the fluid motion associated with vortices into sound waves, creates sufficient entropy that the results are in full compliance with the second law of thermodynamics. Nevertheless, the BEC remains trapped indefinitely in a non-thermal state, with lower entropy than that expected at full thermal equilibrium.  It is an intriguing question whether the Onsager vortex state is an example of a non-thermal fixed point~\cite{Berges2008a}, and if this behaviour persists in the full quantum dynamics of the system.

\begin{acknowledgments}
We thank Eivind Hauge, Ben Powell, Ashton Bradley, Chris Vale, Elena Ostrovskaya, Tamara Davis, and Gerard Milburn for their helpful comments.  We acknowledge financial support from the  Australian Research Council via Discovery Projects DP130102321 (T.S., K.H.) and DP1094025 (M.J.D.).
\end{acknowledgments}

\bibliographystyle{apsrev}
\bibliography{tortises_prl}
\clearpage
\usecounter{figure}
\setcounter{figure}{0}

\renewcommand{\thefigure}{S\arabic{figure}}
\renewcommand{\theequation}{S\arabic{equation}}

\section{Thermodynamics of point vortices}

The thermodynamics of point vortices provides a framework for understanding the behaviour of the vortex gas in the Gross-Pitaevskii dynamics.  
For true point vortices with no core structure, as described by the two-dimensional Coulomb gas \cite{Salzberg1963a,Hauge1971a,Minnhagen1987a}, the critical temperature $T_{\rm PC}$ of the positive temperature PC transition occurs at $\beta_{\rm PC}E_0=1$, where $\beta=1/k_{\rm B}T$ is the inverse temperature and $E_0 =  \rho_s \kappa^2/4\pi $, with $\rho_s$ the (superfluid) density of particles, and $\kappa = h/m$ is the circulation quantum, where $h$ is Planck's constant and $m$ the mass of an atom. For non-zero size vortex cores, the critical temperature of this phase transition shifts \cite{Viecelli1995a} toward the Berezinskii--Kosterlitz--Thouless (BKT) value of $\beta_{\rm BKT}E_0=2$ \cite{Berezinskii1971a,Berezinskii1972a,Kosterlitz1973a,Nelson1977a}. 

\begin{figure*}[!t]
\centering
\includegraphics[width=1.7\columnwidth]{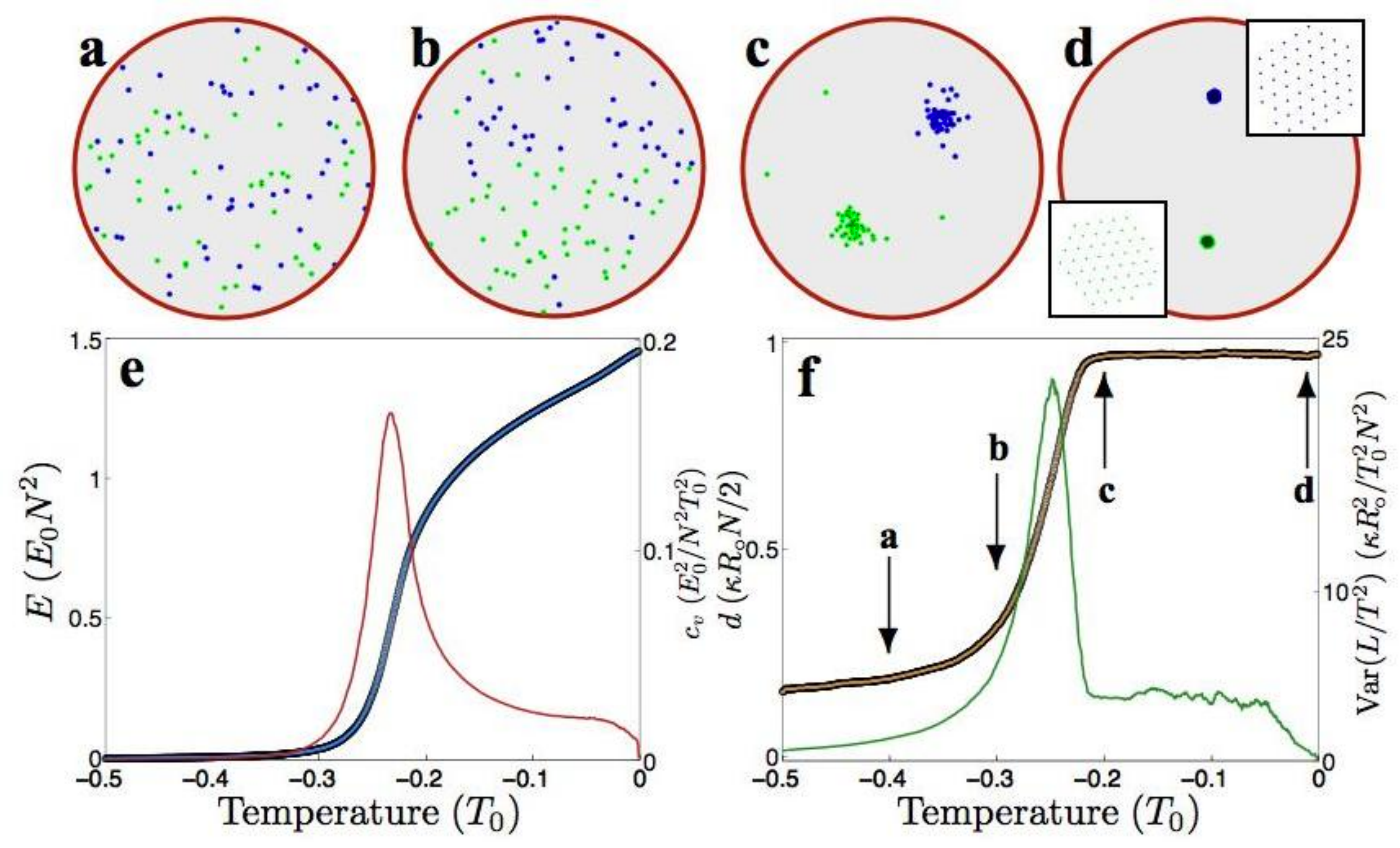}
\caption{(Color online) Onsager vortex phase transition within the point vortex model at negative temperature.  (a--d) Representative vortex configurations corresponding to the  temperatures $T/T_0 = -0.4, -0.3, - 0.2$ and $- 0.001$ respectively. (e) Energy (blue, left axis) and specific heat (red, right axis)  (f) Dipole moment (orange, left axis) and angular momentum fluctuations (green, right axis) as functions of temperature determined by the Monte Carlo calculations. The Onsager vortex phase transition occurs in the vicinity of $T =-T_0/4$. At very low negative temperatures the vortices crystallize, as shown in (d). The specific heat has power law behaviour $c_v \propto |T-T_c|^{-\alpha}$ near the critical point with measured exponents $\alpha \approx 5/2$ and $\alpha ' \approx 3/4$ on the disordered and ordered sides of the transition, respectively. For the limit of zero core point vortices we measure $\alpha\approx\alpha'\approx 5/2 $.\label{S1n}}
\end{figure*}

Similar to the usual free-energy arguments used for estimating the critical temperature of the PC and BKT transitions, the critical temperature $T_{\rm OV}$ for the negative-temperature Onsager vortex phase transition can be predicted \cite{Kraichnan1980a} by estimating the \emph{a priori} probability of the Onsager vortex configuration to be $\left( \xi_{N}/R_\circ \right)^{2N}$, where $\xi_{N}$ is the radius of the vortex cluster placed in a system of linear dimension $R_\circ$. The energy of a multiply quantized vortex dipole with pair separation $R_\circ$ and $N/2$ vortices of each sign is $E=(1/{2\pi}){\rho_s(N\kappa/2)^2}\ln\left( R_\circ /\xi_{N}\right)$.  The product of the \emph{a priori} probability and the Boltzmann factor $e^{-\beta E}$ increases without bound when $\xi_{N}\to 0$ unless $\beta > \beta_c$, where $\beta_c E_0N^2/2+2N=0$. This energy-entropy balancing predicts a negative temperature phase transition to the Onsager vortex state with a vortex number dependent critical temperature $T_{\rm OV}=-\pi N\rho_s\hbar^2/4 m^2 k_{\rm B}$. 
 
In Fig.~\ref{S1n} we show Monte Carlo results for the point vortex model for $T<0$ with $N=100$. Figure~\ref{S1n}(a--d) are representative microstate vortex configurations,  where we observe an abrupt change in the behaviour of the vortices at a critical temperature $T_{\rm OV}$. In Fig.~\ref{S1n}(a), which corresponds to a large negative temperature, high entropy state, the vortex positions are mostly uncorrelated. Figure~\ref{S1n}(b)  is near the critical temperature and begins to display clustering of like-signed vortices. At higher energies, as in Figs.~\ref{S1n}(c--d),  well-defined clusters of like-signed vortices emerge corresponding to the OVs.  At very low negative temperatures, in Fig.~\ref{S1n}(d), we observe vortex clusters crystallised into triangular arrays. This is a consequence of using finite-sized vortex cores. Viecelli \cite{Viecelli1995a} has reported similar results for a positive temperature system with the sign of the vortex-vortex interactions changed. 

The sudden emergence of the ordered OV state is highly suggestive of a phase transition.  Figure~\ref{S1n}(e) shows the energy $H_\circ/N^2$ and specific heat $c_v=\langle H_\circ^2 - \langle H_\circ\rangle^2\rangle/T^2$ as functions of temperature.  The appearance of OVs in the microstates is accompanied by a steep rise in the energy and a corresponding sharply peaked, lambda-shaped feature in the specific heat. Figure~\ref{S1n}(f) shows the vortex dipole moment $d$ and the  angular momentum fluctuations ${\rm var}(L/T^2)=\langle L^2 - \langle L\rangle^2\rangle/T^2$ as functions of temperature, where $L = \sum_i q_i |\mathbf{r}_i|^2$. There is a peak in the fluctuations of the angular momentum near the critical point, and a jump in the vortex dipole moment from uncorrelated to a fully correlated state.  The vortex dipole moment serves as an order parameter for this transition. It is non-zero at infinite temperature due to the finite vortex number. The critical temperature for the transition to the OV state \cite{Onsager1949a,Kraichnan1980a,Viecelli1995a} is $T_{OV} \approx -0.25 \;T_0$, where  $T_0=E_0N/k_{\rm B}$.

\section{Emergence of Onsager vortices in the Gross--Pitaevskii dynamics}

The striking similarity between the observation of vortex clustering in the Gross--Pitaevskii dynamics (Fig. 1) and the $T < 0$ transition  occurring in the point vortex model thermodynamics (Fig.~\ref{S1n}) suggests that our GPE simulations constitute a dynamical progression across the Onsager vortex transition.    Here we connect the dynamics of the turbulent 2D BEC to the thermodynamics of the point-vortex model. 

\begin{figure}
\centering
\includegraphics[width=\columnwidth]{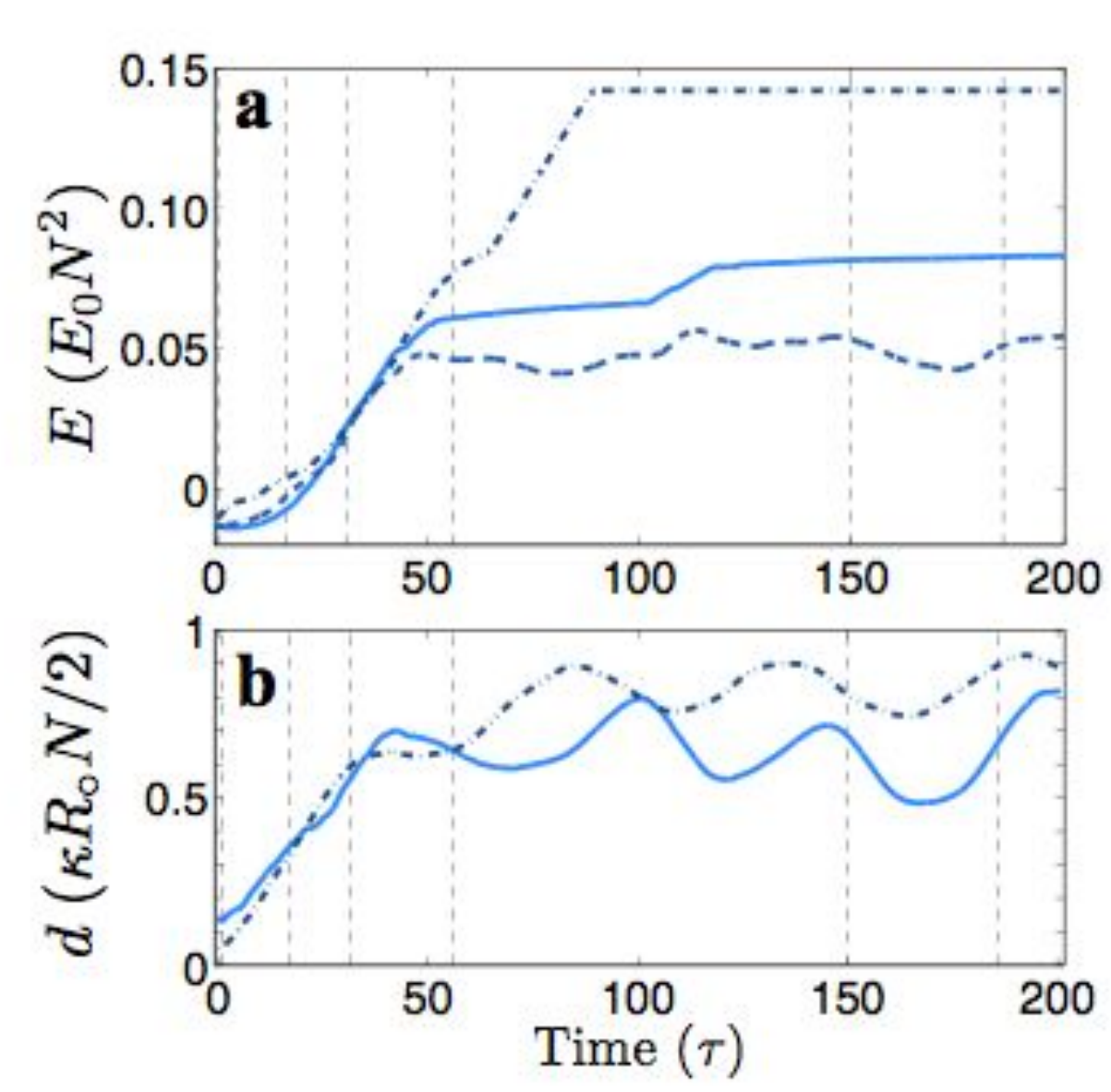}
\caption{(Color online) Comparison of the GPE and point vortex observables. (a) Point vortex model energetics as a function of time.  Blue (dashed): Energy of the point-vortex model corresponding to the location of the GPE vortices.  Light blue (solid): Kinetic energy of the GPE model mapped to the point vortex model.  Dark blue (dashed-dotted):  energy from the point vortex simulations with added vortex annihilation.  All curves cross the critical energy for Onsager vortex formation near $t=40 \tau$. (b) Dipole moment of the GPE model (solid light blue) and point vortex model with vortex annihilation (dashed-dotted dark blue) as functions of time. All curves are sliding averages to instantaneous values over a time window of $15 \tau$.\label{S2n}}
\end{figure}

At each time step of the GPE simulation we determine the locations of the vortex cores in the $z=0$ plane. We calculate the energy of this vortex configuration in the 2D point-vortex model  [see Eq.~(\ref{pointvortex})] The results are shown in Fig.~\ref{S2n}(a), along with the GPE kinetic energy mapped to the point-vortex model. Figure~\ref{S2n}(a) also shows the energy of a sample trajectory of the point-vortex model with vortex-antivortex annihiliation.  Given the significant differences between the models, the evolution of the energies are remarkably similar. The emergence of the Onsager vortices in the GPE simulations at $t\approx 40\;\tau$ [Fig.~1] is consistent with the finite-size system exceeding a threshold energy. From the GPE vortex configuration we calculate the vortex dipole moment $d$, plotted in Fig.~\ref{S2n}(b). At time $t=0$ this is $d\approx 0.2$, corresponding to $T \leq -0.4 \;T_0$. It rapidly rises and reaches the equilibrium critical value at time  $t\approx 40\; \tau$, before oscillating on a long time scale with a period corresponding to the dipole orbits of the OVs~\cite{Neely2010a}. The  dipole moment of the same sample trajectory of the point vortex model is shown, and again is remarkably similar.

\begin{figure*}
\centering
\includegraphics[width=1.8\columnwidth]{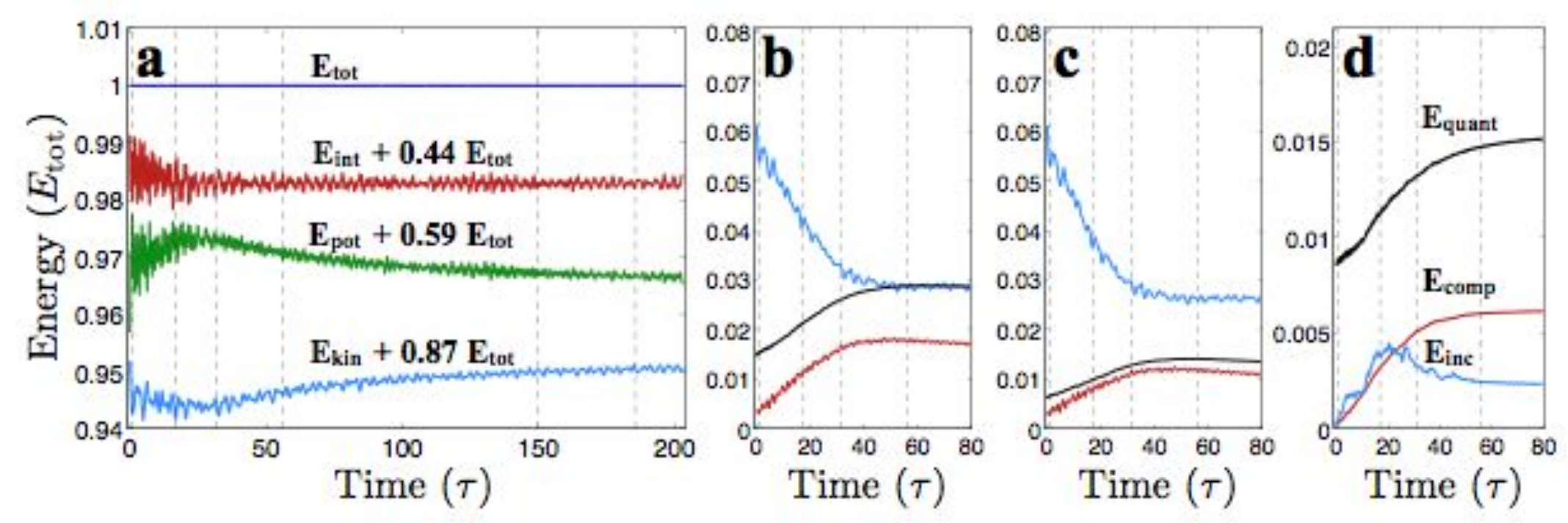}
\caption{(Color online) Gross--Pitaevskii energetics as a function of time. (a) Red: interaction energy;  Green: potential energy;  Light blue: kinetic energy.  Constants have been added to these curves as indicated in the figure so that fluctuations are more visible. The high frequency oscillations in the energies occur at the axial oscillator frequency $\omega$. 
(b) Blue: total incompressible kinetic energy;  Red: total compressible kinetic energy;  Black: total quantum pressure. (c) Blue: transverse incompressible kinetic energy;  Red: transverse compressible kinetic energy;  Black: transverse quantum pressure. (d) Blue: axial incompressible kinetic energy;  Red: axial compressible kinetic energy;  Black: axial quantum pressure. The times corresponding to images Fig.~1(a)--(f) are indicated in all subfigures by the vertical dashed lines.\label{S3n}}
\end{figure*}

In Fig.~\ref{S3n} we present the different energy components of the GPE simulation as a function of time. Despite clear trends in the kinetic, potential and interaction energies in Fig.~\ref{S3n} (a), the emergence of Onsager vortices is not obviously identifiable from this information. Figure~\ref{S3n}(b) shows the kinetic energy of the GPE simulation divided in three components. These are the incompressible (blue) and compressible (red) kinetic energies and the quantum pressure (black) due to the condensate density gradients \cite{pethicksmith}. Figures~\ref{S3n}(c--d) further breaks down the energies shown in ~\ref{S3n}(a) into their radial and axial components, respectively. The key result quantified by these data is the conversion of the incompressible kinetic energy of the vortices into the density fluctuations (compressible sound waves and quantum pressure) by the vortex evaporative heating mechanism. Once the OVs have been formed, all energies become relatively settled. We associate the hump in the axial incompressible kinetic energy in Fig.~\ref{S3n}(d) with the excitation of the low energy axial Kelvin waves \cite{Thomson1880a,Pitaevskii1961a,Dalibard2003a,Fetter2004a} along the vortex lines by the annihilation events. These axial excitations subsequently decay by emitting sound waves \cite{Simula2008a,Simula2013a}.

\section{Spectral properties}
 
We briefly mention the spectral properties of the Onsager vortex state in the GPE simulation.
For driven-dissipative 2D turbulence, Kraichnan predicted the transport of energy from small scales to ever larger ones via a scale invariant, inverse energy cascade mechanism~\cite{Kraichnan1967a}.  This has an incompressible kinetic energy spectral distribution $E_i(k)=C \epsilon^{2/3} k^{-5/3}$, where $C$ is a constant, $\epsilon$ is the rate of energy transfer and $k$ the wavenumber for the excitation.  The $-5/3$ exponent is the same as the celebrated Kolmogorov scaling law for three-dimensional turbulence \cite{Kraichnan1980a}. In classical systems, theory predicts that the additional conservation of enstrophy (integral of the squared vorticity) leads to a direct enstrophy cascade with $E_i(k)=C'\eta^{2/3} k^{-3}$, less a logarithmic correction, where $C'$ is a constant and $\eta$ is an enstrophy transfer rate \cite{Kraichnan1967a}.  However, as the GPE simulations are conservative, the dual cascade scenario of the Kraichnan model, which assumes energy injection and dissipation, is strictly not applicable to our system.

\begin{figure}
\centering
\includegraphics[width=\columnwidth]{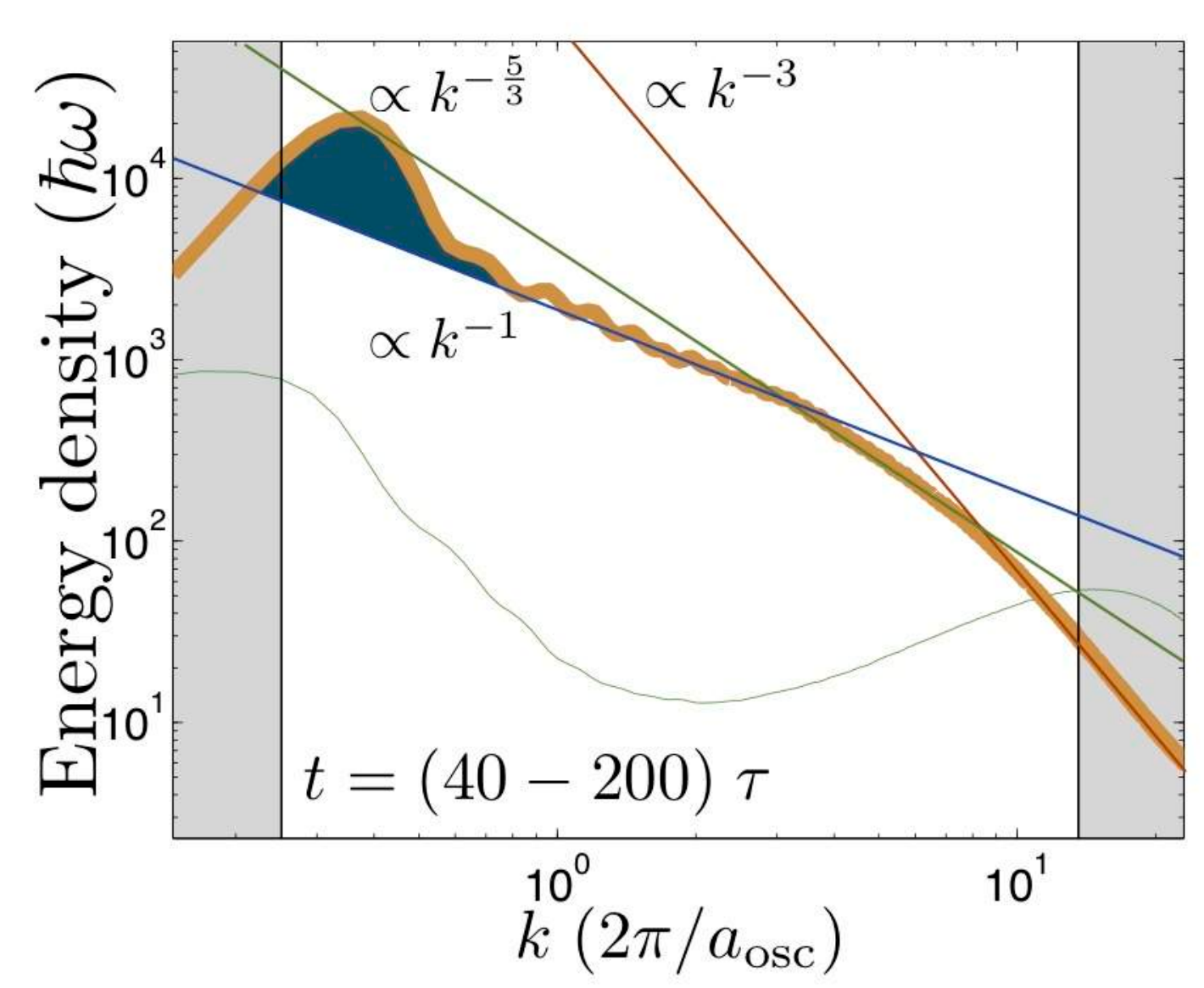}
\caption{Spectral features of the GPE wave function following the emergence of Onsager vortices.
Compressible (green) and incompressible (orange) kinetic energies, quantum pressure inclusive, measured from the $z=0$, $r<R_0$ slice of the GPE wave function and averaged over time $t/\tau \in [40,200]$. The hump that appears at the longest wavelengths, which also appears in the energy spectra of the point vortex model velocity field for the same vortex configurations (not shown), is due to the EBC of OVs. The straight lines are guides to data.\label{S4}}
\end{figure}

Figure \ref{S4} shows the spectral distributions of the incompressible (orange) and compressible (green) kinetic energies as a function of wavenumber measured for the circular plane $r<R_0,z=0$ of the GPE wave function following the emergence of the OVs.  While the short wavelength behaviour with $E_i(k)\propto k^{-3}$ matches that of Kraichnan's direct enstrophy cascade scenario, this spectral power law is not related to the configuration of vortices but is instead determined by the scaling behaviour expected for the Fourier transform of an inverse power-law spatial velocity field around an isolated quantized vortex core \cite{Krstulovic2010a,Bradley2012a}. In two dimensions the long wavelength spectrum with $E_i(k)\propto k^{-1}$ arises from the far-field  of isolated vortices in two dimensions~\cite{Krstulovic2010a,Bradley2012a} and uncorrelated vortex configurations \cite{Kusumura2013a}. The shaded long-wave feature highlights the spectral condensate due to the Einstein--Bose condensation \cite{Kraichnan1967a} of Onsager vortices. Recently, Billam \emph{et al}. \cite{Billam2013a} simulated the dynamics of a Bose-Einstein condensate using a GPE in a doubly-periodic two-dimensional domain with periodic arrays of vortex clusters as initial states and observed dynamical rearrangement of the seed clusters into larger Onsager vortices together with an accompanying spectral condensation signature in the incompressible kinetic energy spectrum, see also \cite{Yatsuyanagi2005a,Yatsuyanagi2009a}.

\section{Connection with classical field methodology}
The Gross-Pitaevskii model is a mean-field approximation for the dynamics of  a Bose-Einstein condensate.  Often it is assumed to be valid near zero temperature when the condensate fraction is close to unity~\cite{pethicksmith}. However, the so-called classical field method introduces a methodology that effectively redeploys the Gross-Pitaevskii equation to account for the effects of quantum and thermal fluctuations~\cite{Blakie2008b}. It has previously been shown, for example, that populating randomised incoherent excitations that subsequently undergo conservative dynamics leads to the  thermalisation of the system~\cite{Davis2001b,Davis2002a,Davis2003a}. 

In contrast, here we observe what appears to be partial thermalization following the addition of randomised \emph{coherent} topological excitations to the BEC in the form of quantized vortices.  Further elaborating the connection between Onsager vortex formation and thermalisation of quasiparticle excitations with the classical field methodology, in particular in the disc-shaped trap geometry, will be the subjects of future investigation.


\section{Gross--Pitaevskii model}

We model the dynamics of the condensate with the three-dimensional Gross--Pitaevskii equation 
\begin{equation}
i\hbar\partial_t\psi({\bf r})=\left[-\frac{\hbar^2}{2m}\nabla^2  + U_{\rm disc}({\bf r})  +g |\psi({\bf r})|^2 \right]\psi({\bf r}), 
\end{equation}
where $g=4\pi\hbar^2 a/ m$, and $a$ is the s-wave scattering length.  The field $\psi({\bf r})$ is normalised to the total number of atoms $N_a$.  We consider a cylindrically symmetric potential
\begin{equation}
U_{\rm disc}({\bf r})= U_0[\tanh(\chi (r-1.12R_\circ )/a_{\rm osc} )+1] + m\omega^2z^2/2,
\end{equation}
so that the atoms are confined radially in a potential that is practically constant out to a radius of $R_\circ $, before steep walls rise.  In the axial $z$-direction  the system is  tightly confined by a harmonic potential of frequency $\omega$ such that the vortices are constrained to essentially two-dimensional motion.  However, the bursts of sound energy released in vortex-antivortex annihilation   results in occasional Kelvin wave excitations of the remaining vortices, as can be seen from the spiralling motion of some vortices in Movie S1.

 The trap parameters of the GPE simulation in Fig.~1 of the main text are $U_0 = 16\;\hbar\omega$, $R_\circ=28\;a_{\rm osc}$, and $\chi=0.6$.  The dimensionless nonlinearity is $gN_a/(a_{\rm osc}^3 \hbar\omega) = \sqrt{125} \times 10^4$.  Assuming $^{87}$Rb atoms with a scattering length of $a = 5.29$ nm and $\omega = 2\pi \times 90$ Hz as in the recent experiment by Neely \emph{et al.}~\cite{Neely2012a},  we find that $N_a \sim 2 \times 10^6$, which is in close correspondence to the number of atoms in that experiment.  The radius of the system is then $R_\circ =71 \;\mu$m.
 
To prepare the initial state for the GPE simulation we begin with the ground state of the system.  We subsequently  numerically imprint the velocity field of an equal number of vortices and antivortices (100 in total) at random locations while ensuring the BEC has near zero angular momentum.  A short period of evolution in imaginary time heals the structure of the vortex cores, and leads to the annihilation of typically 20 vortices of opposite signs, thus leaving 80 vortices in the initial state.  Imprinting a larger number of vortices results in more vortices  in the emergent Onsager vortex clusters. 

In Fig.~\ref{S3n} we compare the 3D Gross--Pitaevskii kinetic energy with that of the point-vortex model.
We measure the 3D Gross--Pitaevskii energy in units of $[E_{\rm GP}]=N_a\hbar\omega$ whilst the 3D energy of the point vortices is in units of $[E_{\rm PV}] = \rho_s\kappa^2  L _e/4\pi$, where $L_e$ is the effective length of the vortices. The scale factor between these is 
\begin{equation}
\eta = \frac{[E_{\rm PV}]}{[E_{\rm GP}]} = \frac{\rho_s\kappa^2  L _e/4\pi}{N_a\hbar\omega} = \frac{\pi \mu L_e}{g' a_{\rm osc} \hbar \omega},
\end{equation}
where $\mu$ is the chemical potential  of the GPE wave function. The point vortex model energy $E_{PV}$ corresponding to a 3D GPE wave function with kinetic energy $E^{\rm KE}_{\rm GP}$ is
\begin{equation}
{E}_{\rm PV}= \eta (E^{\rm KE}_{\rm GP} - E^{\rm KE}_0),
\end{equation}
where  $E^{KE}_0$ is the kinetic energy of the GPE ground state. The point vortex energies are computed on a disc of unit radius and therefore energy shift of $\nu=1.8326 \frac{\rho_s\kappa^2}{4 \pi}N$ is subtracted from ${E}_{\rm PV}$ to account for the physical radius of the disc.

\section{Point vortex Hamiltonian}
The model Hamiltonian we use to describe point vortices confined to a two-dimensional disc of unit radius  is \cite{PointinLundgren1976a,Viecelli1995a,Buhler2002a,Yatsuyanagi2005a}
\begin{eqnarray}
H_\circ &=& H_{\rm vortices} + H_{\rm images} + H_{\rm self},\notag\\
&=& -\frac{\rho_s\kappa^2}{4 \pi}\sum_{i<j} s_is_j\log\left(r^2_{ij}\right) \notag\\
&+& \frac{\rho_s\kappa^2}{4 \pi}\sum_{i<j} s_i s_j\log\left(1  -2(x_i x_j+ y_iy_j)+ |z_i|^2|z_j|^2\right) \notag\\
&+& \frac{\rho_s\kappa^2}{4 \pi}\sum_i s_i^2\log\left(1- r^2_{i}\right),
\label{pointvortex}
\end{eqnarray}
where $x_j={\rm Re}(z_j)$ and $y_j={\rm Im}(z_j)$ are the 2D Cartesian coordinates of the $j$th vortex of winding number $s_j=\pm1$ measured in units of $R_\circ$.  The term $H_{\rm vortices}$ accounts for the pair-wise logarithmic long-range Coulomb interaction between the vortices separated by a distance $r_{ij} = \sqrt{(x_i-x_j)^2+(y_i-y_j)^2}$. For vortices of the same sign, $s_is_j>0$, the  interaction is repulsive, whereas for vortices with opposite circulations, $s_i s_j<0$, it is attractive. The second term $H_{\rm images}$ arises due to the boundary --- it represents the interaction of the system vortices with the images of other vortices.   The last term $H_{\rm self}$ describes the interaction of the system vortices with their own image.

\section{Point vortex thermodynamics}
To calculate the thermodynamic properties of the point vortex gas, we employ a Markov chain Monte-Carlo method using a Metropolis algorithm to sample the vortex configurations at different temperatures \cite{Viecelli1995a}. We impose a hard core vortex radius of $\xi$  to account for the finite size of vortex cores in physical systems.
We have performed temperature sweeps for $T\in(0, + \infty)$ and for $T\in(-\infty,0)$, sampling
 $5\times 10^6$ microstates at each temperature after a burn-in of $10^6$ steps. The parameters used for the results presented in Fig.~\ref{S1n}, which show typical configurations of point vortices as a function of temperature, are $N=100$ and $R_\circ /\xi=100$. 
 
\section{Point vortex dynamics with pair annihiliation}
The point vortex Hamiltonian~Eq.~\ref{pointvortex} describes a dynamical system with the equation of motion
\[
i \kappa s_j \frac{\partial z_j}{\partial t} = \frac{\partial H_\circ}{\partial z_j}. 
\]
Integrating the equation of motion for an initial configuration in the normal state of the phase diagram results in chaotic motion of the vortices, with the constants of motion being the energy $H_\circ$, angular momentum $L_\circ=\kappa \sum_j s_j|z_j|^2$, and the total number of vortices $N$ and their circulations.

To mimic the vortex-antivortex pair annihilation that occurs in the GPE simulation, our numerical algorithm removes all pairs that are driven by the dynamics to within an ``annihilation distance" $d_a=0.08\;R_\circ$ from one another.  This choice of $d_a$ results in the mean number of vortices at long times agreeing with the GPE simulations.  Smaller values of $d_a$ result in fewer annihilation events.

We have simulated the dynamics of an ensemble of 500 stochastically sampled initial vortex  configurations (see Fig.~3 in the main text.) The initial states are obtained by placing  vortices in the system at random locations one at a time.  A vortex is only added to the system if it does not fall within the annihilation distance of any existing vortices, until a total of 80 are present.  This self-avoiding placement mimics the GPE initial conditions, whereby the short imaginary time integration in the preparation causes like pairs that are close to annihilate, or unlike pairs to repel one another.  The initial state is only used for the ensemble if the resulting configuration has an energy close to the initial energy of the point vortex energy of the GPE simulation. Movie S2 shows an example of the point vortex dynamics for a single trajectory of the ensemble.

\section{Supplemental Movies}
\subsection{Movie S1---Simulation of the Gross-Pitaevskii model}
Simulation of the dynamics of the Bose-Einstein condensate imprinted with vortices as described in text using the Gross-Pitaevskii model. The movie shows the evolution of the condensate column density. The locations of the vortices and antivortices are shown using blue and green markers, respectively, and their numbers are shown in the top right corner. Time is shown in the top left corner. The white circle has a radius of $R_0$. The dipole moment vector is displayed as a straight white line.

\subsection{Movie S2---Simulation of the point-vortex model}
Simulation of the evaporative heating of point vortices with the addition of vortex annihilation. The locations of the vortices and antivortices are shown using blue and green markers, respectively, and their numbers are shown in the top right corner. Time is shown in the top left corner. The black circle has a radius of $R_0$. The dipole moment vector is displayed as a straight black line.

\end{document}